# From probabilistic mechanics to quantum theory

## U. Klein

Springer

CHAPMAN UNIVERSITY | INSTITUTE FOR QUANTUM STUDIES

**REGULAR PAPER**

# From probabilistic mechanics to quantum theory

**U. Klein**

© The Author(s) 2019

**Abstract** We show that quantum theory (QT) is a substructure of classical probabilistic physics. The central quantity of the classical theory is Hamilton's function, which determines canonical equations, a corresponding flow, and a Liouville equation for a probability density. We extend this theory in two respects: (1) The same structure is defined for arbitrary observables. Thus, we have all of the above entities generated not only by Hamilton's function but also by every observable. (2) We introduce for each observable a phase space function representing the classical action. This is a redundant quantity in a classical context but indispensable for the transition to QT. The basic equations of the resulting theory take a "quantum-like" form, which allows for a simple derivation of QT by means of a projection to configuration space reported previously [Quantum Stud Math Found 5:219–227, 2018]. We obtain the most important relations of QT, namely the form of operators, Schrödinger's equation, eigenvalue equations, commutation relations, expectation values, and Born's rule. Implications for the interpretation of QT are discussed, as well as an alternative projection method allowing for a derivation of spin.

## 1 Introduction

General agreement regarding the meaning of the quantum-mechanical formalism has not been achieved so far. This lack of clarity is closely related to a lack of clarity regarding the relation between quantum theory (QT) and classical physics. The present-day ideas on the quantum–classical interface have been established more than 90 years ago. They have never undergone a critical reexamination, despite the fact that a wealth of new information, both experimentally and theoretically, has been obtained since then.

Einstein's claim that QT must be an ensemble theory, and not a theory about individual particles, is neither generally accepted nor has it ever been refuted. Although not a majority view, it has been supported by many outstanding physicists. An excellent review article about this "statistical interpretation" or "ensemble interpretation"

U. Klein (✉)
Institute for Theoretical Physics, University of Linz, 4040 Linz, Austria
e-mail: ulf.klein@jku.at

is available [1]. Among the earliest, papers written in the spirit of Einstein's interpretation are pioneering works by VanVleck [2], Bopp [3,4], and Schiller [5].

Since then, several other papers sharing this interpretation, but using a variety of different methods and assumptions, have been published. An incomplete list includes attempts to understand QT in terms of ensembles either in phase space [6–10] or in configuration space [11–17]. These works clarify several important aspects of QT. There is, however, still considerable room for improvements, as regards the number and nature of the assumptions used to derive QT.

The second main opinion, in particular promoted by Bohr in discussions with Einstein [18], claims that QT is a "complete" theory for individual particles. This "individuality interpretation" is more common than the ensemble interpretation despite the obvious fact that QT makes probabilistic predictions (this fundamental discrepancy is the origin of all the ongoing discussions). The prevailing opinion seems to be that the question has already been decided and one should not call into doubt the chosen path.

To understand QT means to clarify its relation to classical physics. But what exactly means "classical physics" ? If we accept Bohr's individuality interpretation, we will identify "classical physics" with classical mechanics, the classical theory of individual particles. On the other hand, if we prefer Einstein's ensemble point of view, we will identify "classical physics" with a probabilistic theory of classical particles. Our objects to study are then statistical ensembles of particles rather than individual particles.

The present paper is the second in a series of works where Einstein's point of view is taken and verified. In the first paper [19], referred to as I, some essential relations of QT were derived using Koopman–von Neumann theory as a starting point. In the present paper, the problem of the derivation of QT is attacked in a more systematic way. In the following Sect. 2, we discuss the quantum–classical interface, taking into account all major theoretical results in this field. This analysis leads to the conclusion that QT should be derivable from a probabilistic version of classical mechanics, a conclusion to be verified in the remaining part of this paper.

Our starting point is the theory of ensembles of classical particles, formulated first (in Sect. 3) with particle coordinates as independent variables, and rewritten then (in Sect. 4) in terms of the more familiar field-theoretic formulation, where coordinates denote points in "space". In our case, "space" means $2n$-dimensional phase space, with coordinates $p$ and $q$ (instead of the $n$-dimensional configuration space with coordinates $q$ used for example in QT). We introduce two basic dynamical variables, the probability density $\rho(q, p, t)$, and the action $S(q, p, t)$, for the probabilistic description of particles moving in the course of time $t$. These variables obey two basic differential equations, the Liouville equation and an action equation. This structure is determined by the definition of a single-phase space function, the Hamiltonian $H(q, p)$. The solutions of the canonical equations define a family of transformations, also referred to as flow, on phase space.

This construction may be formulated for arbitrary observables $A(q, p)$, with corresponding independent parameter $\alpha$; it is known that each $A(q, p)$ generates a one-parameter Lie group, realized as a subgroup of canonical transformations [20]. This extension is reported in Sect. 5. Thus, each $A(q, p)$ defines a flow for varying $\alpha$ and two dynamical variables $\rho_A(q, p, \alpha)$, $S_A(q, p, \alpha)$ obeying corresponding differential equations. We call the resulting theory, where a multitude of fields occur and modern probability theory plays an important role the "Hamilton–Liouville–Lie–Kolmogorov theory" (HLLK). Introducing for each $A$ a complex-valued classical state variable, one obtains, after some manipulations, a theory which shares many structural properties with QT (see Sects. 7–10). Certain phase space operators $\hat{L}_A$ (introduced already in I), which are generalizations of Koopman–Neumann operators, represent the HLLK counterpart of quantum operators $\hat{A}$.

This work is based on the fundamental assumption that all physical fields must be formulated in configuration space; as a consequence, we have to perform a projection of phase space to configuration space. If the quantum-like form of HLLK, reported in Sect. 8, is used, the transition from to QT becomes very simple. This transition, which is performed in Sect. 11, creates Schrödinger's equation, the form of the quantum-mechanical operators, commutator relations, expectation values, and Born's rule. Thus, all fundamental structural properties of QT may be derived from HLLK. A general discussion of the present approach is given in Sect. 12, concluding remarks are made in the final Sect. 13.

## 2 The quantum–classical interface

Here, we compare the mathematical structures of QT and classical physics. More precisely, we compare the mathematical structures of QT and two classical theories, namely: the deterministic description of classical particles (classical mechanics), and the probabilistic description of classical particles (probabilistic mechanics). We are interested, in particular, in the question, which one of these three theories can be considered as a "covering theory", in the sense that it can be reduced to another theory. Correspondingly, an important mathematical tool of this investigation is counting degrees of freedom.

Let us start with a comparison of the mathematical structures of classical mechanics and QT. We restrict ourselves to mechanical systems which may be cast in the Hamiltonian form of classical dynamics. This formulation is also best suited for comparison with QT. The canonical equations are given by the following:

$$\dot{q} = \frac{\partial H(q, p)}{\partial p}, \quad \dot{p} = -\frac{\partial H(q, p)}{\partial q}, \tag{1}$$

where $H(q, p)$ is Hamiltons's function (here assumed time-independent) and $q_k, \ p_k \ (k = 1, \ldots, n)$ are the components of the generalized coordinates $q$ and conjugate momenta $p$, respectively. The dot denotes a derivative with respect to $t$. The state of the system, at each instant of time $t$, is given by a point in phase space determined by the $2n$ real numbers $q, \ p$. The basic law of QT is Schrödinger's equation:

$$-\frac{\hbar}{\imath} \frac{\partial}{\partial t} \psi(q, t) = \hat{H} \psi(q, t). \tag{2}$$

Here, the Hamilton operator is defined by $\hat{H} = H(q, \frac{\hbar}{\imath} \frac{\partial}{\partial q})$ and the dynamic variable $\psi$ depends, at a particular time $t$, on the $n$ coordinates $q_1, \ldots, q_n$ of the classical configuration space. The state of a system in QT, at a particular time, is given by a point in Hilbert space, specified by the real and imaginary parts of the complex-valued function $\psi(q, t)$. The number of degrees of freedom required to specify a function is of course uncountable infinite. Let us for comparison introduce a symbol for a large number, say $\mathbb{I}$, to denote any finite approximation for the uncountable infinite number of degrees of freedom associated with a real axis (at the times of Sophus Lie, we would use the symbol $\infty$). A precise definition is not required; in particular, the precise mathematical definition of the cardinality of the continuum is completely useless for our purpose. All we have to know is that $\mathbb{I} \gg n$. Using this symbol, we may say that a system described in classical mechanics by $2n$ numbers requires $2\mathbb{I}^n$ numbers for its description in QT.

Thus, there is a gigantic mismatch between classical mechanics and QT as regards the number of degrees of freedom of one and the same physical system. Classical mechanics cannot, of course, be the covering theory of QT, but the inverse may be true. In fact, according to the prevailing opinion, expressed by authorities [21], QT reduces in the classical limit $\hbar \to 0$ to classical mechanics. This is, however, not the case. It has been shown recently that almost all states of QT do not reduce to corresponding solutions of classical mechanics [22]. There is still the possibility left, that classical mechanics is the classical counterpart of QT in the sense that it can be used to derive the form of all quantum-mechanical operators. This possibility of a "quantization", in the sense a of a complete structural similarity, must also be excluded, as Groenewold reported already many years ago his no-go theorem for quantization, which says that a consistent map between the structures of classical and quantum-mechanical observables does not exist [23].

Let us consider now the second possibility that the classical counterpart of QT is not classical mechanics, but entails "only" the probabilistic description of classical particles. In this case, the fundamental dynamical variable is given by the probability density $\rho(q, p, t)$, which has to obey Liouville's equation:

$$\frac{\partial \rho}{\partial t} + \frac{\partial \rho}{\partial q} \frac{\partial H}{\partial p} - \frac{\partial \rho}{\partial p} \frac{\partial H}{\partial q} = 0. \tag{3}$$

The basic law takes now the form of a partial differential equation (as in QT), because an infinite number of degrees of freedom is required to describe the behavior of particles. It is linear (as in QT), which means that almost all Cauchy data will lead to well-behaved solutions.

The probabilistic theory described by Eq. (3) has a simple conceptual structure; the probabilistic element stems from incomplete knowledge of initial values, while the deterministic behavior, realized by trajectories in phase space, of classical particles remains intact. In statistical mechanics, the number of degrees of freedom is, in contrast to probabilistic mechanics, much larger than the number of conserved quantities. As a consequence, statistical mechanics does not make probabilistic predictions, as probabilistic mechanics does.

A state of this theory is now, in analogy to QT, a function on phase space (similar to a point in Koopman–Neumann Hilbert space [24]). Using our above estimate, the number of degrees of freedom is $2\mathbb{I}^{2n}$ (the additional factor of 2 stems from the fact that a second dynamical variable will be introduced in Sect. 4). The set of all states contains as a limiting case "pure states" described by Delta-function-like initial conditions. Despite certain similarities, probabilistic mechanics and classical mechanics are completely different theories, in particular with regard to their relation to QT.

If we compare now the number of degrees of freedom of probabilistic mechanics and QT, we find again a gigantic mismatch, but this time the number of reals $2\mathbb{I}^{2n}$ required to describe a classical probabilistic state *exceeds* the corresponding number $2\mathbb{I}^{n}$ of QT by a vast amount. This result, which is at first sight surprising, implies that QT does not reduce to probabilistic mechanics in the limit $\hbar \to 0$. To discuss the mathematical relation of both theories, the Wigner–Weyl formulation of QT is most appropriate [25,26]. The Wigner function, defined in terms of $\psi(q, t)$ and $\psi(q', t)$, is given by $W(Q, P, t) = W_\hbar(Q, P, t)$, where

$$W_w(Q, P, t) = \frac{1}{(2\pi w)^n} \int dr \, \psi^\star \left( Q - \frac{r}{2}, t \right) \psi \left( Q + \frac{r}{2}, t \right) \exp \left\{ -\frac{\iota}{w} Pr \right\}, \tag{4}$$

and $Q = \frac{q+q'}{2}$, $r = -q + q'$. Note that the quantity $\hbar$ is part of the definition of $W(Q, P, t)$. The equation of motion of Wigner's function, first obtained by Moyal [27], differs from Liouville's equation by an infinite number of terms which all seem to vanish if $\hbar \to 0$. However, the $\hbar-$dependence of $W(Q, P, t)$ must also be taken into account. If this is done, it turns out that the time-evolution of $W(Q, P, t)$, for almost all potentials (the exceptions are the same as for the classical limit of Schrödinger's Eq. [22]) in the limit $\hbar \to 0$ does *not* follow Liouville's Eq. [7,28]. Thus, in agreement with our estimate, QT does not reduce to probabilistic mechanics in the classical limit. The simple reason is that the dimension of configuration space is only half the dimension of phase space. Consequently, $Q$, $P$ are not independent variables and the "phase space of QT" is a "mock phase space" [29].

As QT does neither reduce to classical mechanics nor to probabilistic mechanics, we have to conclude that probabilistic mechanics, or an appropriate extension of this theory, can somehow be reduced to QT. This conclusion is astonishing at the first sight. We know that QT is the theory actually realized in nature and it is, at least in this sense, certainly superior to any classical theory. One would, therefore, expect (according to our reductionistic way of thinking) that QT embraces at least one of the classical theories and reduces to it in the classical limit. However, this is not the case. Probabilistic mechanics is superior to QT in the sense that it contains a higher number of degrees of freedom and it is this number which rules the mathematical relation between both theories, not the question which theory is realized in nature. The old idea that the quantum constant $\hbar$ plays the role of an accuracy limit in phase space [4] is in perfect agreement with our present hierarchical ordering of theories in terms of degrees of freedom. The considerations of this section may be summarized in the form of the following working hypothesis: "QT is a configuration space version of the probabilistic description of classical particles".

## 3 Lagrangian phase space ensembles defined by the Hamiltonian function

The systems that we want to study are not individual trajectories but ensembles of trajectories. It is important to distinguish clearly between these two kinds of systems. We start our construction of HLLK by defining an ensemble as an infinite set of individual trajectories, which differ from each other by their states $q_0$, $p_0$ at a fixed time $t_0$. We study deterministic phase space ensembles, which means that each trajectory is completely known if a single point on it is known. We start with the fundamental "Lagrangian" description of a classical ensemble, where the positions of the individual members are determined by the trajectory label $q_0$, $p_0$ and the time $t$ [30]. The transition to the more common "Eulerian" description will be performed in the next section.

The solutions of the equations of motion (1) and their inverses are written, suppressing the dependence on $t_0$, as

$$q = Q(q_0, \, p_0, \, t), \quad p = P(q_0, \, p_0, \, t), \tag{5}$$

$$q_0 = Q_0(q, \, p, \, t), \quad p_0 = P_0(q, \, p, \, t). \tag{6}$$

These relations are subject to the constraints $q_0 = Q(q_0, \, p_0, \, t_0)$, $p_0 = P(q_0, \, p_0, \, t_0)$, which means that the $n-$component quantities $q_0$, $p_0$ are the particle coordinates and momenta at the initial time $t_0$. A (deterministic) phase space ensemble [31] is the set of all possible trajectories:

$$\mathbb{E}_{DP} = \left\{ q = Q(q_0, \, p_0, \, t), \, p = P(q_0, \, p_0, \, t) \, | \, (t, q_0, p_0) \in \mathbb{R} \times \mathbb{R}^n \times \mathbb{R}^n \right\}. \tag{7}$$

This system possesses an infinite (continuous) number of degrees of freedom. The independent variables are members of the set $\{t, \, q_0, \, p_0 \, | \, (t, q_0, p_0) \in \mathbb{R} \times \mathbb{R}^n \times \mathbb{R}^n\}$. Correspondingly, the basic equations of our ensemble theory are not given by (1) but by an infinite number of equations of the form (1), but each one with *prescribed* initial values $q_0$, $p_0$. We may write

$$\left\{ \dot{q} = \frac{\partial H(q, p, t; q_0, q_0)}{\partial p}, \quad \dot{p} = -\frac{\partial H(q, p, t; q_0, q_0)}{\partial q} \, \middle| \, (q_0, p_0) \in \mathbb{R}^n \times \mathbb{R}^n \right\}, \tag{8}$$

to denote this infinite set. The independent variables $q_0$, $p_0$ must be taken into account even if our basic equations do not contain any derivatives with respect to these variables (It is for this reason that the present field theory does not take the standard form of a partial differential equation but the "degenerate" form of an infinite number of ordinary differential equations). A Lagrangian (here, we use the standard meaning of this term from classical mechanics) for our basic equations, not written down here, contains an integral over all contributions from the initial values $q_0$, $p_0$.

In a classical probabilistic theory, particles move according to deterministic laws, but deterministic predictions are nevertheless impossible, because the particle's trajectories cannot be localized with certainty. This uncertainty can be quantitatively described by introducing a Lagrangian probability density for the states of the particles at a particular time $t_0$. Such probabilistic theories, where trajectories still exist, have been classified as "type 2 theories" in a recent work of the present author [15].

Denoting the Lagrangian probability density at time $t_0$ by $\rho_0(q_0, p_0)$, the infinitesimal quantity $\rho_0(q_0, p_0) \, \mathrm{d}^n q_0 \, \mathrm{d}^n p_0$ gives the probability to find a particle within an infinitesimal volume element located at $q_0$, $p_0$. The corresponding Lagrangian probability density for the same particle (labeled by $q_0$, $p_0$) at a later time $t$ is denoted by $\rho_L(q_0, p_0, t)$. The probability to find this particle at time $t$ within an infinitesimal volume element located at $q(t)$, $p(t)$ is given by $\rho_L(q_0, p_0, t) \, \mathrm{d}^n q(t) \, \mathrm{d}^n p(t)$. The fact that trajectories remain intact in the course of time implies the relation:

$$\rho_L(q_0, \, p_0, \, t) \, \mathrm{d}^n q(t) \, \mathrm{d}^n p(t) = \rho_0(q_0, \, p_0) \, \mathrm{d}^n q_0 \, \mathrm{d}^n p_0. \tag{9}$$

Note that $q_0 = q(t_0)$, $p_0 = p(t_0)$ and $\rho_0(q_0, p_0) = \rho_L(q_0, p_0, t_0)$. Eq. (9) indicates that the mapping between $q_0$, $p_0$ and $q(t)$, $p(t)$, defined by the solutions (5) of equations of motion (1), determines the relation between $\rho_0(q_0, p_0)$ and $\rho_L(q_0, p_0, t)$. This mapping is denoted as Hamiltonian (or symplectic) flow. Indeed, performing a transformation of variables, one finds that

$$\rho_L(q_0, \, p_0, \, t) J(t) = \rho_0(q_0, \, p_0), \tag{10}$$

where the Jacobian $J(t)$ is the determinant of the Jacobian matrix $\mathcal{D}$ associated with the mapping (5),

$$J(t) = det\mathcal{D} = \frac{\partial(Q_1, \ldots, Q_n, P_1, \ldots P_n)}{\partial(q_{0,1}, \ldots, q_{0,n}, p_{0,1}, \ldots p_{0,n})}. \tag{11}$$

A mathematical fact of fundamental importance is the uniqueness of the solutions of the first-order system (1) for specified initial values $q_0, p_0$. This fact implies $J(t) > 0$ and the invertibility of the mapping (5). This is also a necessary condition for the global validity of the solutions (5), which we assume to hold true. Actually, the stronger statement $J(t) = 1$ holds true and this fact will be used below in a number of transformations.

We may now calculate the observable output of our Lagrangian ensemble theory. The expectation value $\bar{A}(t)$ of the (Lagrangian) observable $A_0(q_0, p_0, t)$ in the (Lagrangian) state $\rho_0(q_0, p_0)$ is given by integrating the product $A_0\rho_0$ over all phase space points $q_0, p_0$:

$$\bar{A}(t) = \int A_0(q_0, p_0, t)\rho_0(q_0, p_0)\mathrm{d}^n q_0\, \mathrm{d}^n p_0. \tag{12}$$

Note that in the Lagrangian formulation, observables depend on time, while states are time-independent.

The Lagrangian formulation outlined here provides the most natural framework for the definition of classical deterministic ensembles. However, the use of particle labels as independent variables is inconvenient and a transition to the more common "Eulerian" description, where points of "space" are used as independent variables, is useful—this is all the more true as we are finally interested in QT, which is a Eulerian (configuration space) theory.

## 4 Eulerian phase space ensembles defined by the Hamiltonian function

The quantities $q$, $p$ and $q_0$, $p_0$ appearing in Eq. (5) denote particle properties. Today, physicists believe that it makes sense to speak about abstract spaces which exist no matter whether particles are present or not. The points of these abstract spaces may be obtained from the Lagrangian coordinates $q_0, p_0$ by performing at each instant of time a transformation from the initial values $q_0, p_0$ to the final (at time $t$) values $q$, $p$ of the particle properties. This uncountable set of transformations:

$$\Phi_t^H : \Omega \to \Omega, \tag{13}$$

is also referred to as *flow*. We introduced here the symbol $\Omega$ to denote $2n-$dimensional phase space, $\Omega = \mathbb{R}_q^n \times \mathbb{R}_p^n$. The flow $\Phi_t^H$ is defined by the solutions (5) of the equation of motion (1) and is in the present context denoted as Lagrangian–Eulerian map [30,32]. The totality of all image points represents the same continuum at each instant of time $t$. This property is responsible for the independent existence of the abstract space created this way. This process replaces the independent coordinates $q_0, p_0$ by new Eulerian coordinates, which are also denoted by $q$, $p$, but characterize now points of (phase) space.

It is remarkable that the Lagrangian–Eulerian map defines a concept of space in terms of particle properties; a careful analysis might possibly lead to the conclusion that it is unnecessary to introduce absolute space with the help of extra axioms.

To perform the Lagrangian–Eulerian map for the basic equations of our ensemble theory, we start from the solutions (5) and the associated inverse Eq. (6). For each Lagrangian function, say $G_L(q_0, p_0, t)$, a corresponding Eulerian function $G(q, p, t)$ is defined by the following:

$$G(q, p, t) = G_L(Q_0(q, p, t), P_0(q, p, t), t), \tag{14}$$

Using the fact that the initial values $q_0$, $p_0$ (considered as a function both of time $t$ and the time-dependent solutions $q$, $p$) do not depend on $t$, a partial differential equation for the Eulerian observable $G(q, p, t)$ may be derived by total differentiation of (14) with respect to $t$. The result takes the form:

$$\frac{\partial G}{\partial t} + \frac{\partial G}{\partial q} \frac{\partial H}{\partial p} - \frac{\partial G}{\partial p} \frac{\partial H}{\partial q} = \frac{\partial G_L}{\partial t}. \tag{15}$$

This is the general Eulerian partial differential equation which represents the most convenient way to study classical ensembles; there is no way to describe individual particles anymore. To characterize a system, appropriate functions on phase space, obtained from corresponding Lagrangian functions, have to be specified.

The most important Eulerian function is the probability density $\rho(q, p, t)$, obtained from the time-independent Lagrangian quantity $\rho_0(q_0, p_0)$. This field may be used to describe the "state" of our (Eulerian) probabilistic ensemble. Then, the right-hand side of Eq. (15) vanishes and we obtain the Liouville equation:

$$\frac{\partial \rho}{\partial t} + \frac{\partial \rho}{\partial q} \frac{\partial H}{\partial p} - \frac{\partial \rho}{\partial p} \frac{\partial H}{\partial q} = 0. \tag{16}$$

All time-independent Lagrangian functions lead to the same differential Eq. (16). As a consequence, this homogeneous linear partial differential equation has a huge manifold of solutions. The Liouville equation, together with the Eulerian expression:

$$\bar{A}(t) = \int A(q, p)\rho(q, p, t)\mathrm{d}^n q \, \mathrm{d}^n p, \tag{17}$$

represent the basic building block of HLLK. The expectation value (17) is obtained from (12) by means of a transformation of variables defined by (6). In contrast to the Lagrangian formulation, Eulerian states depend on time, while Eulerian observables are time-independent (here, we find analogies with different formulations of time-dependence in QT, see, e.g., [20]).

The above equations are sufficient to calculate most quantities of interest of HLLK, at least as far as the observable $H(q, p)$, ruling the time-dependence of observables, is concerned. There is, however, a certain arbitrariness in the definition of a "state" of HLLK, even if we restrict ourselves to the single observable $H(q, p)$ (see below). This arbitrariness can be used to facilitate the transition to QT. We introduce a second Eulerian dynamical variable, describing the purely deterministic part of the evolution in phase space. As an appropriate quantity, we consider the classical action

$$S = \int_{t_0}^{t} \mathrm{d}t' L\left(q(t'), \dot{q}(t')\right), \tag{18}$$

where $q(t')$ are the real paths, i.e., the solutions of the equations of motion. At these stationary points, the integrand may be written in the form:

$$L(q, \dot{q}) = \bar{L}(q, p) = p \frac{\partial H(q, p)}{\partial p} - H(q, p), \tag{19}$$

where $q$, $p$ are solutions of the canonical Eq. (1). If we identify the $2n$ integration constants with the initial coordinates and momenta $q_0$, $p_0$ at $t_0$, then we have a Lagrangian function $S_{t_0}(q_0, p_0, t)$ depending explicitly on time $t$. The transition to the corresponding Eulerian differential equation may be performed with the help of Eq. (15), as reported in detail in I. We find that our second Eulerian dynamical variable $S(q, p, t)$ has to obey the equation:

$$\frac{\partial S}{\partial t} + \frac{\partial S}{\partial q}\frac{\partial H}{\partial p} - \frac{\partial S}{\partial p}\frac{\partial H}{\partial q} = \bar{L}, \tag{20}$$

where $\bar{L} = \bar{L}(q, p)$ is defined as in Eq. (19). Equations (16) and (20) for $\rho(q, p, t)$ and $S(q, p, t)$, together with Eq. (17) for the expectation values $\bar{A}(t)$, represent an extended set of basic equations of HLLK.

The action equation represents the purely deterministic part of HLLK, as shown by the fact that $\rho(q, p, t)$ does not occur in (20). This kind of decoupling between $S(q, p, t)$ and $\rho(q, p, t)$ is the most important feature of classical probabilistic physics. We introduced the new variable $S(q, p, t)$, which might seem redundant at the first sight, to formulate a theory where this decoupling can break down. In fact, in QT, this decoupling *will* break down and this is the deeper reason why we introduced the field $S(q, p, t)$.

We note in passing that Eq. (20) may be projected to configuration space by neglecting the $p-$dependence of $S$ and by replacing $p$ by the gradient of $S$. In this way, one obtains the Hamilton–Jacobi equation in a very quick way.

## 5 Eulerian phase space ensembles defined by arbitrary observables

The above concepts were all defined in terms of a single-phase space function, the Hamiltonian $H(q, p)$, which is the most important observable. We required nowhere a special functional form of $H(q, p)$. Thus, all these concepts may also be defined for an arbitrary phase space function $A(q, p)$. We incorporate these additional structures, for arbitrary $A(q, p)$, in the framework of HLLK, anticipating that it will be useful from a physical point of view.

Instead of defining canonical equations for a "Hamiltonian" $A(q, p)$, let us start from the regions in phase space where $A(q, p)$ takes a constant value, say $a$. These regions are given by the level sets:

$$\mathbb{L}_A(a) = \{(q, p) \in \Omega \mid A(q, p) = a \in \mathbb{R}_A\}, \tag{21}$$

where the number $a$ belongs to the image $\mathbb{R}_A \subseteq \mathbb{R}$ of $A(q, p)$. Let us next consider a connected level set of the form:

$$\mathbb{L}_A(a \mid q_0, p_0) = \{(q(\alpha), p(\alpha)) \in \Omega \mid A(q(\alpha), p(\alpha)) = a \in \mathbb{R}_A, \ (q_0, p_0) = (q(0), p(0))\}, \tag{22}$$

which defines a curve parameterized by the real parameter $\alpha$ and crossing the point $(q_0, p_0)$ at $\alpha = 0$. As a consequence of the fact that $a$ is a constant, the points in $\mathbb{L}_A(a \mid q_0, p_0)$ obey the condition:

$$\frac{\mathrm{d}}{\mathrm{d}\alpha} A(q(\alpha), p(\alpha)) = \frac{\partial A}{\partial q_k}\dot{q}_k + \frac{\partial A}{\partial p_k}\dot{p}_k = 0, \tag{23}$$

where the dot denotes now derivation with respect to $\alpha$. Condition (23) holds true if $q(\alpha)$, $p(\alpha)$ are solutions of the ordinary differential equations:

$$\dot{q}_k = \frac{\partial A(q, p)}{\partial p_k}, \quad \dot{p}_k = -\frac{\partial A(q, p)}{\partial q_k}, \tag{24}$$

with boundary conditions $q(0) = q_0, \ p(0) = p_0$. Obviously, Eq. (24) agree with (1) if $A$ and $\alpha$ are identified with the Hamiltonian $H$ and the time $t$, respectively. The solutions of (24) and their inverses are denoted as follows:

$$q = Q^A(q_0, p_0, \alpha), \quad p = P^A(q_0, p_0, \alpha), \tag{25}$$
$$q_0 = Q_0^A(q, p, \alpha), \quad p_0 = P_0^A(q, p, \alpha). \tag{26}$$

Thus, an observable $A(q, p)$ takes a constant value along an integral curve of the canonical Eq. (24) defined by the same observable. The dimension of the independent variable $\alpha$, which labels the points on the integral curves, is the dimension of $\hbar$ divided by the dimension of $A$; observables related this way are said to be conjugate, or complementary, to each other.

The relations (24) play an important role in Lie's theory of transformation groups and in the theory of Hamiltonian vector fields. The solutions define again, in analogy to (13), a family of mappings:

$$\Phi^A_\alpha : \Omega \to \Omega \tag{27}$$

of phase space onto itself. The fact that such a mapping exists may be expressed, using the language of level sets, by means of the relation: $\Omega = \left\{ (Q^A(q_0, p_0, \alpha), P^A(q_0, p_0, \alpha)) | (q_0, p_0) \in \Omega \right\}$, which must be true for arbitrary $\alpha \in \mathbb{R}$. The integral curves of all equations of the form (24) cover for each value of the parameters $\alpha$ the whole of phase space. This statement is, of course, only true if the solutions of (24) are well defined throughout phase space [32]).

Let us consider as an example the flows created by the fundamental phase space observables $q_1, \ldots, p_n$ (other examples may be found, e.g., in [33]). For $A = q_i$, where $i$ is a fixed integer from the set $\{1, \ldots, n\}$, the parameter, say $\pi$, labeling the integral curves of (24) has the dimension of momentum. The solutions do not depend on $\pi$ with the exception of $p_i(\pi) = -\pi + p_i^0$. Thus, the integral curves are lines parallel to the $p_i$−axis; the observable $q_i$ creates translations, with negative sign, in the direction of the corresponding component $p_i$ of the momentum. For $A = p_i$ the parameter, say $\chi$, has the dimension of position. The solutions do not depend on $\chi$ with the exception of $q_i(\chi) = \chi + q_i^0$; the observable $p_i$ creates translations in the direction of $q_i$. More generally, the observables $A(q, p)$ play the role of generators of one-parameter groups of (canonical) transformations on phase space [20].

We may associate, in analogy to Sect. 4, two Eulerian phase space functions with each observable $A(q, p)$, a probability density $\rho_A(q, p, \alpha)$ and an action function $S_A(q, p, \alpha)$, which obey the differential equations:

$$\frac{\partial \rho_A}{\partial \alpha} + \frac{\partial \rho_A}{\partial q} \frac{\partial A}{\partial p} - \frac{\partial \rho_A}{\partial p} \frac{\partial A}{\partial q} = 0, \tag{28}$$

$$\frac{\partial S_A}{\partial \alpha} + \frac{\partial S_A}{\partial q} \frac{\partial A}{\partial p} - \frac{\partial S_A}{\partial p} \frac{\partial A}{\partial q} = \bar{L}_A. \tag{29}$$

The inhomogeneous term in (29) is the Lagrange function $\bar{L}_A$ associated with $A$, which is defined by the following:

$$\bar{L}_A(q, p) = p \frac{\partial A(q, p)}{\partial p} - A(q, p), \tag{30}$$

and the action expressed in Lagrangian coordinates, which is used to derive (29), is given by the following:

$$S_A(q_0, p_0, \alpha) = \int_{\alpha_0}^{\alpha} d\alpha' \left[ P(q_0, p_0, \alpha') \frac{\partial Q(q_0, p_0, \alpha')}{\partial \alpha'} - A(Q(q_0, p_0, \alpha'), P(q_0, p_0, \alpha')) \right]. \tag{31}$$

The basic equations of the probabilistic theory defined by $A(q, p)$ are mathematically equivalent to the relations reported in the last section. On the other hand, the physical meaning of the various theories, obtained for different $A(q, p)$, is quite different.

## 6 Comparison with standard probability theory

In this section, we discuss the relation between HLLK and modern probability theory; To do this, we have to use $\rho$ as state variable, leaving a possible role of $S$ aside. We start by recalling the basic features of time-independent (Kolmogorovian) probability theory. Then, we give a short review of the extension to time-dependent phenomena. Finally, we discuss how HLLK fits into the time-dependent framework.

### 6.1 Time-independent probability theory

The result of a single experiment, called "trial", is denoted as "outcome" (or elementary event) and is mathematically represented by a point $\omega \in \Omega$. An "event" is a subset $E \subseteq \Omega$. The set $\mathcal{F}$ of admissible subsets must be a $\sigma$-algebra: this means, among other things, that $\mathcal{F}$ is closed under formation of complements and countable unions [34]. The remaining essential element is a universal probability measure $P(E)$, a function defined on $\mathcal{F}$ and normalized according to $P(\Omega) = 1$. The measures take the form of integrals over $E \in \mathcal{F}$:

$$P(E) = \int_E \mathrm{d}\omega \, \rho(\omega), \tag{32}$$

assigning weight to the different points of $E$ according to a probability density $\rho(\omega) \geq 0$. A classical probability space is given by the triple $(\Omega, \mathcal{F}, P)$. Classical observables are arbitrary (sufficiently smooth) functions $A : \Omega \to \mathbb{R}$, $\omega \mapsto A(\omega)$, which are in a probabilistic context referred to as random variables.

The most important number characterizing the statistical behavior of an observable $A(\omega)$ is its expectation value $\bar{A}_\rho$, in the state $\rho$, which is defined by the following:

$$\bar{A}_\rho = \int \mathrm{d}\omega \, A(\omega)\rho(\omega). \tag{33}$$

Most experimental predictions of a statistical theory may be written in the form (33). For example, $P(E)$ as given by (32) takes the form of (33) if $A(\omega)$ is identified with the characteristic function $I_E(\omega)$ of the set $E$, defined by the following:

$$I_E(\omega) = \begin{cases} 1 & : \omega \in E \\ 0 & : \omega \notin E. \end{cases}$$

Similarly, the average of an observable $A$ with respect to a subset $E$ is given by (33) if $A(\omega)$ is replaced by $A(\omega)I_E(\omega)$.

The mathematical structure $(\Omega, \mathcal{F}, P)$ can be interpreted as theoretical image of a physical experiment, which is of a purely probabilistic (time-independent) nature. The space $\Omega$ and the functional form of $\rho(\omega)$ have to be chosen in such a way that the experimental setup is properly described. This framework, where no free parameter like the time $t$ appears, is of course not appropriate for a description of HLLK. It can, however, be used to describe HLLK at a single instant of time, the initial time $t_0$, when the functional form of $\rho(\omega)$ is completely under our control. Thus, we may identify $\rho(\omega)$ with the initial distribution $\rho_0(\omega)$ (and $\Omega$ with the phase space). This means, we start at $t_0$ from a *Lagrangian* description of particle motion; the coordinates $\omega$ describe properties of particles and not abstract space points. The expectation value (33) agrees, after an appropriate change in notation, with the Lagrangian expression (12) taken at $t = t_0$.

### 6.2 Time-dependent probability theory

To take the time-dependence (replacing here for definiteness the general parameter $\alpha$ by the familiar time $t$) of real systems into account, the probability space $(\Omega, \mathcal{F}, P)$ must be supplemented by the continuous set of transformations $\Phi_t : \Omega \Rightarrow \Omega$, $\omega_0 \mapsto \omega = \phi_t(\omega_0)$, $\omega_0, \omega \in \Omega$, which is provided by the solutions (25) of the equations of motion. This set, say $\Phi$, of transformations on $\Omega$ represents a dynamical system. The extended structure $(\Omega, P, \Phi)$ (we omit here the symbol $\mathcal{F}$ for brevity) may either be interpreted as a dynamical probability space, or as a probabilistic dynamical system (in the mathematical literature, it is denoted as measure-theoretic dynamical system). From a physical point of view, a dynamical system $(\Omega, \Phi)$ alone, representing the dynamics of an uncountable number of individual systems, does not make sense; it must be supplemented by a (probability) density describing the distribution of individual systems in phase space.

Using this extended framework, the expectation value (33) must be replaced by the time-dependent Lagrangian expression (12), which, in the present notation, takes the form:

$$\bar{A}_\rho(t) = \int d\omega_0 A_0(\omega_0, t) \rho_0(\omega_0). \tag{34}$$

Performing a change of variables from $\omega_0$ to $\omega$, as given by the dynamical map $\omega = \phi_t(\omega_0)$, we obtain the corresponding Eulerian expression:

$$\bar{A}_\rho(t) = \int d\omega A(\omega) \rho(\omega, t), \tag{35}$$

which represents the most common way to write the expectation value. To define a measure in the time-dependent setting, we introduce the symbol $E_t$ to denote a particular subset of $\Omega$ chosen at time $t$. The time-dependent measure is then defined by the following:

$$P(E_t, t) = \int_{E_t} d\omega \, \rho(\omega, t). \tag{36}$$

This expression may again be derived from the time-independent Lagrangian expression (32) by means of a transformation $\omega = \phi_t(\omega_0)$. This derivation shows also that $P(E_t, t) = P(E_{t'}, t')$, i.e., the measure defined by (36) is invariant under the dynamical map; the term "measure-preserving transformation" is frequently used.

The mathematical structure $(\Omega, P, \Phi)$ represents the theoretical image of an experiment associated with $A$ which shows both probabilistic and deterministic aspects. The probabilistic aspect is due to uncertainty at a initial time $t_0$, the deterministic aspect belongs to the subsequent movement of the ensemble in phase space. We are interested in dynamical systems of Hamiltonian type, with Hamiltonian $A$. Then, the probability density remains constant on curves (solutions of the canonical equations defined by $A$) of constant $A$.

### 6.3 Comparison of time-dependent probability theory with HLLK

The mathematical structure, $(\Omega, P, \Phi)$, where the dynamical system $\Phi$ is of Hamiltonian type, represents the basic building block of HLLK, but does not exhaust it. Comparing both structures, we note the following points:

– HLLK may contain several versions of this building block; each observable $A(\omega)$ may be used to define a structure $(\Omega, P_A, \Phi_A)$. Thus, HLLK contains several components which describe independent experiments, but share the same sample space $\Omega$. Using the standard notation, HLLK may be written in the form $(\Omega; P_A, \Phi_A; P_B, \Phi_B, \ldots)$.

– Each component of HLLK is implicitly defined by two functions which may be chosen freely, namely the observable $A(\omega)$ and the initial value $\rho_A^0(\omega)$ of the probability density. Indeed, the set of Hamiltonian transformations $\Phi_A$ is implicitly defined by the form of $A(\omega)$; the same is true for the time-dependence of $P_A$ and $\rho_A(\omega, t)$. However, the initial value $\rho_A^0(\omega)$ of $\rho_A(\omega, t)$ may be chosen freely. If we denote the pair $A$, $\rho_A^0$ by the symbol $E_A$, HLLK may also be written in the form $HLLK = [\Omega, E_A, E_B, \ldots]$.

– Each component $E_A$ belonging to HLLK is associated not only with an observable $A(\omega)$ but also with a corresponding independent variable $\alpha$. Of course, we assume that Hamilton's function $H(\omega)$, with its associated parameter $t$, is always contained in the set of relevant observables. Comparing with standard time-dependent probability theory, we note that HLLK, in its present general form, has a multi-parameter structure which calls for a reduction.

In its general form, HLLK describes several (dynamical) experimental arrangements each one associated with a particular observable (a reduction of the dynamics will be performed in Sect. 8). More precisely, we assume that the experimental arrangement defined by $A$ is constructed in such a way that sharp values of the state variable $\rho_A$ may be obtained. The various experimental arrangements are in principle independent from each other, as regards the measurement of the various probability densities. However, there may be an interplay between different observables, in the sense that expectation values of arbitrary observables (random variables) may be measured in each one of these experiments. The question arises if experimental arrangements exist, where state variables $\rho_A$, $\rho_B$ belonging to different observables, may be measured simultaneously. A related question will be studied in Sect. 9.

## 7 Introducing a complex state variable

The relations (28) and (29) represent the basic equations of our general theory, defined by an arbitrary observable $A(q, p)$. The first of these describes the deterministic evolution, for varying $\alpha$, of the probabilistic quantity $\rho_A$; the second describes the evolution of the deterministic quantity $S_A$. There is a certain freedom as regards the choice of a "state variable" for our system. The quantity $\rho_A$ (or a function thereof) is certainly the first choice; in this case, we have the single basic law (28), and the second basic Eq. (29) plays no role. On the other hand, a proper combination of $\rho_A$ and $S_A$ is also allowed; the action $S_A$ describes the same physics as $\rho_A$, as least far as the deterministic evolution in phase space is concerned. In this section, we introduce a combination of $\rho_A$ and $S_A$ which will be useful for the transition to QT.

Let us first introduce a more compact notation. Points $q_1, \ldots, q_n, p_1, \ldots, p_n$ in $\Omega$ will be denoted by $\omega$ if no need for distinction between $q$ and $p$ arises. The Poisson bracket of two phase space functions $A(\omega)$, $B(\omega)$ is defined by the following:

$$\{A, B\} = \sum_{k=1}^{n} \left( \frac{\partial A}{\partial q_k} \frac{\partial B}{\partial p_k} - \frac{\partial A}{\partial p_k} \frac{\partial B}{\partial q_k} \right). \tag{37}$$

Besides the obvious properties of bilinearity and antisymmetry, the Poisson bracket obeys the relations:

$$\{AB, C\} = A\{B, C\} + \{A, C\} B, \tag{38}$$

$$\{\{A, B\}, C\} + \{\{B, C\}, A\} + \{\{C, A\}, B\} = 0, \tag{39}$$

which are referred to as product rule and Jacobi identity [20]. Two observables with vanishing Poisson bracket are said to be in involution.

Using this notion, the basic equations (24), (28), and (29) take the form:

$$\dot{q}_k = \{q_k, A\}, \quad \dot{p}_k = \{p_k, A\}, \tag{40}$$

$$\frac{\partial \sqrt{\rho_A}}{\partial \alpha} + \{\sqrt{\rho_A}, A\} = 0, \tag{41}$$

$$\frac{\partial S_A}{\partial \alpha} + \{S_A, A\} = \bar{L}_A. \tag{42}$$

Denoting the points of phase space by $\omega$ the solutions of (40) may be written in the compact form $\omega = \phi_\alpha(\omega_0)$, with the inverse given by $\omega_0 = \phi_\alpha^{-1}(\omega)$. This uncountable set of transformations $\Omega \Rightarrow \Omega$, with Jacobian equal to unity, represents a one-parameter Lie group with unit element $\phi_{\alpha_0}$.

Given two arbitrary phase space functions $F(\omega)$, $A(\omega)$, the condition $\{F, A\} = 0$ is equivalent to the statement that $F$ is a constant of motion with regard to the canonical equations (40) defined by $A$; this statement remains true if the roles of $F$ and $A$ are exchanged. The Poisson bracket $\{A, B\}$ may also be written in the form $\hat{D}_A B$, where $\hat{D}_A$ is a linear differential operator defined by the following:

$$\hat{D}_A \cdot = \{A, \cdot\} = \sum_{k=1}^n \left( \frac{\partial A}{\partial q_k} \frac{\partial \cdot}{\partial p_k} - \frac{\partial A}{\partial p_k} \frac{\partial \cdot}{\partial q_k} \right). \tag{43}$$

It is referred to as Lie derivative with respect to $A$. Application of $\hat{D}_A$ means differentiation along the integral curves defined by the observable $A$. As $\hat{D}_A B = -\hat{D}_B A$ holds, a vanishing Poisson bracket between two observables means that each one is invariant under the flow generated by the other observable.

Using the Lie derivative, the partial differential Eqs. (41) and (42) take the form:

$$\left( \frac{\partial}{\partial \alpha} - \hat{D}_A \right) \sqrt{\rho_A} = 0, \tag{44}$$

$$\left( \frac{\partial}{\partial \alpha} - \hat{D}_A \right) S_A = \bar{L}_A. \tag{45}$$

In (41) and (44), we replaced the dynamical variable $\rho_A$ by $\sqrt{\rho_A}$; this is allowed, since an arbitrary function of a solution of the Liouville equation is again a solution. Thus, the *state* of the ensemble associated with the observable $A$ is defined to be $\sqrt{\rho_A}$. This definition is preferable as far as the transition to QT, reported in Sect. 11, is concerned. In a purely classical context, both state definitions, $\rho_A$ and $\sqrt{\rho_A}$, are equivalent and the simpler term $\rho_A$ is sometimes more convenient.

If we want to construct a new state variable, which is a function of both $\rho_A$ and $S_A$ and obeys a single (linear) evolution equation, we are more or less automatically led to the range of complex numbers. It is easy to see that the two decoupled Eqs. (44) and (45) may be written in the form:

$$\left( \frac{\hbar}{\imath} \frac{\partial}{\partial \alpha} + \hat{L}_A \right) \phi_A = 0, \qquad \hat{L}_A = -\frac{\hbar}{\imath} \hat{D}_A - \bar{L}_A, \tag{46}$$

if a new complex-valued state variable $\phi_A$, defined by the following:

$$\phi_A(\omega, \alpha) = \sqrt{\rho_A(\omega, \alpha)} \exp \frac{\imath}{\hbar} S_A(\omega, \alpha), \tag{47}$$

is introduced. Thus, the basic equations of HLLK may be written in a form resembling Schrödinger's equation. There are certain conditions to be fulfilled [35], to guarantee the uniqueness of the "classical wave function" $\phi_A$, which are not important in the present context.

A natural expression for an inner product is obtained by extending the standard definition of QT from configuration space to phase space:

$$(\phi, \psi) = \int d\omega \phi^*(\omega) \psi(\omega). \tag{48}$$

The operator $\hat{L}_A$ is self-adjoint with regard to this inner product. Equation (46) is a generalization of the Koopman–von Neumann equation (which is obtained for $\bar{L}_A = 0$) as discussed in more detail in I. These changes in notation make the equations look slightly more "quantum-like", but are irrelevant as far as the physical content of the present classical theory is concerned. However, they will turn out to be very useful later, when the transition from HLLK to QT is performed.

## 8 Reduction to a single-parameter theory

In its present general form, HLLK has a multi-observable (multi-parameter) structure. Besides the Hamiltonian $H$, and its associated independent parameter $t$, other observables $A$, with associated parameters $\alpha$, appear in the theory. The physical meaning of these "non-time-like" components of HLLK is still unclear.

To clarify this point, let us consider, as an example, a single particle (set $n = 3$) and identify the observable $A$ with the $z$−component of the angular momentum, $A(q, p) = L_3 = q_1 p_2 - q_2 p_1$. The parameter $\alpha$ is dimensionless and can be identified with an angle of rotation. The solutions of the canonical Eqs. (40) are given by the following:

$$q_1 = q_1^0 \cos \alpha - q_2^0 \sin \alpha, \quad q_2 = q_1^0 \sin \alpha + q_1^0 \cos \alpha, \quad q_3 = q_3^0 \tag{49}$$

$$p_1 = p_1^0 \cos \alpha - p_2^0 \sin \alpha, \quad p_2 = p_1^0 \sin \alpha + p_1^0 \cos \alpha, \quad p_3 = p_3^0. \tag{50}$$

These relations may be interpreted as kinematical relations in $\mathbb{R}^6$, describing the change of the coordinates of a point under a rotation with an angle $\alpha$. They may also be interpreted as a family of mapping $\mathbb{R}^6 \to \mathbb{R}^6$. Let us consider next the state function $\rho_{L_3}(q, p, \alpha)$ corresponding to this flow. If the initial values at $\alpha = 0$ are distributed according to a probability density $\rho_{L_3}^0(q^0, p^0)$, and if the transformation inverse to Eqs. (49) and (50) is denoted by $q_i^0 = Q_i^0(q, p, \alpha)$, $p_i^0 = P_i^0(q, p, \alpha)$, then the probability density $\rho_{L_3}(q, p, \alpha)$ is given by the following:

$$\rho_{L_3}(q, p, \alpha) = \rho_{L_3}^0(Q_i^0(q, p, \alpha), P_i^0(q, p, \alpha)).$$

This is the solution of Eq. (41) which agrees with $\rho_{L_3}^0(q^0, p^0)$ at $\alpha = 0$. For arbitrary initial values $\rho_{L_3}^0(q^0, p^0)$, the Eulerian function $\rho_{L_3}(q, p, \alpha)$ will generally depend on $\alpha$. However, if $\rho_{L_3}^0(q^0, p^0)$ is chosen to be invariant under the mapping defined by $A$, then $\rho_{L_3}(q, p, \alpha)$ cannot depend on $\alpha$. This happens if $\rho_{L_3}^0(q^0, p^0)$ depends only on $L_3$ or on observables in involution with $L_3$. For example, if $\rho_{L_3}^0$ depends on the invariants $(q_1^0)^2 + (q_2^0)^2$, $q_3^0$ and $(p_1^0)^2 + (p_2^0)^2$, $p_3^0$, then $\rho(q, p, \alpha) = \rho^0\left(q_1^2 + q_2^2, q_3, p_1^2 + p_2^2, p_3\right)$ is also invariant under the mapping defined by $A$, does not depend on $\alpha$, and its Poisson bracket with $A$ vanishes according to (41).

Let us come back now, with this example in mind, to our question concerning the physical meaning of the parameters $\alpha$ and the fields $\rho_A(\omega, \alpha)$. A "multi-parameter theory", taking the dependence of all densities $\rho_A(\omega, \alpha)$ on the corresponding parameters $\alpha$ into account, does probably not correspond to anything realized in nature. Therefore, we restrict ourselves to the common "single-parameter structure", where the dynamics is only taken into account for $A = H$ and $\alpha = t$. For brevity, the letter $H$ in $\rho_H(\omega, t)$ will be omitted and the term "Liouville equation", without any addition, will refer to the Liouville equation defined by $H$. For all other observables $A \neq H$ and $\alpha \neq t$ only the "stationary" solutions of (41), not depending on $\alpha$, will be taken into account. This seems to be the only natural way to reduce the general form of HLLK to a standard dynamical theory (actually of the same type as QT) while taking into account the individual role of observables to the greatest possible extent.

Using the notation, where the state variable is given by $\sqrt{\rho_A}$, the basic equations for these "stationary" fields, say $\bar{\rho}_A(q, p)$, are given by the following:

$$\left\{ \sqrt{\bar{\rho}_A}, A \right\} = 0. \tag{51}$$

The vanishing of the left-hand side means that $\bar{\rho}_A$ is a constant of motion with regard to the canonical equations (40). There is a particularly close connection between $\bar{\rho}_A$ and $A$; both the values of $\bar{\rho}_A$ and of $A$ are constant along the solutions of (40). In an analogous way, one obtains $\alpha$−independent action fields $\bar{S}_A(q, p)$ as solutions of:

$$-a + \left\{ \bar{S}_A, A \right\} = \bar{L}_A, \tag{52}$$

where $a$ is a real number of the same dimension as $A$. The relation between $S_A(q, p, \alpha)$ and $\bar{S}_A(q, p)$ is given by $S_A(q, p, \alpha) = -a\alpha + \bar{S}_A(q, p)$; a corresponding decomposition for $\rho_A(q, p, \alpha)$ is forbidden by the normalization condition of probability.

The two basic stationary Eqs. (51) and (52) may again, in close analogy to Sect. 7, be rewritten as a single relation if a proper complex state variable $\bar{\phi}_A$, not depending on $\alpha$, is introduced. The result takes the form of an eigenvalue equation:

$$\hat{L}_A \bar{\phi}_A = a \bar{\phi}_A. \tag{53}$$

The eigenvalues $a$ must be real, because the operators $\hat{L}_A$ are self-adjoint. In terms of $\bar{\phi}_A$ and $\bar{S}_A$, the stationary state variable $\bar{\phi}_A$ is given by the following:

$$\bar{\phi}_A(\omega) = \sqrt{\bar{\rho}_A(\omega)} \exp \frac{\iota}{\hbar} \bar{S}_A(\omega). \tag{54}$$

The eigenvalue equation (53) may also be derived from the $\alpha-$dependent Schrödinger Eq. (46), using the method of separation of variables. To obtain "quantum-like" relations like (53), the original formalism had to be extended, by introducing a new variable $S_A$ and defining a new state variable. While the physical content of the extended theory is essentially the same as before, it offers new mathematical methods to solve the classical equations of motion.

After this reduction process, we have now a version of HLLK which contains two classes of observables. The first class is that of dynamical observables and it contains only a single element, namely Hamilton's function. Its associated parameter is the time $t$, and its associated fields, state variables, and operators will be denoted by $\rho(q, p, t)$, $S(q, p, t)$, $\phi$, $\hat{L}$, omitting the index $H$. These fields are obtained by solving the classical time-dependent Schrödinger equation (this notation will be justified later):

$$\left( \frac{\hbar}{\iota} \frac{\partial}{\partial t} + \hat{L} \right) \phi = 0. \tag{55}$$

The second class is that of "stationary" observables. It contains all other observables $A$, $B$, .. with parameters $\alpha, \beta, \ldots$, and fields $\bar{\rho}_A(q, p)$, $\bar{S}_A(q, p)$, $\bar{\phi}_A$; $\bar{\rho}_B(q, p)$, $\bar{S}_B(q, p)$, $\bar{\phi}_B$; etc, which are determined by eigenvalue equations:

$$\hat{L}_A \bar{\phi}_A = a \bar{\phi}_A, \quad \hat{L}_B \bar{\phi}_B = b \bar{\phi}_B, \quad \text{etc}, \tag{56}$$

not depending on $\alpha, \beta, \ldots$. Of course, a stationary equation of the type (56) exists also for the Hamiltonian $H$; it is obtained by solving (55) by the method of separation of variables.

After this reduction of our multi-parameter problem, we have now a situation analogous to QT. In fact, a Schrödinger equation exists in QT only for a single operator, the Hamiltonian $\hat{H}$, while all other operators $\hat{A}, \hat{B}, \ldots$ are characterized by eigenvalue equations not depending on the corresponding parameters $\alpha, \beta, \ldots$.

## 9 Simultaneous measurements of several observables

Only the case of an experimental arrangement associated with a single variable $A$ was discussed up to now. We may ask under which circumstances the *simultaneous* measurement of states $\bar{\rho}_A$, $\bar{S}_A$ (or $\bar{\phi}_A$) and $\bar{\rho}_B$, $\bar{S}_B$ (or $\bar{\phi}_B$), corresponding to two different observables $A$ and $B$, may be performed. The theoretical description of such simultaneous measurements requires that phase space functions $\bar{\rho}_{AB}$, $\bar{S}_{AB}$ exist, which obey the conditions:

$$\left\{ \sqrt{\bar{\rho}_{AB}}, A \right\} = 0, \quad -a + \left\{ \bar{S}_{AB}, A \right\} = \bar{L}_A \tag{57}$$

$$\left\{ \sqrt{\bar{\rho}_{AB}}, B \right\} = 0, \quad -b + \left\{ \bar{S}_{AB}, B \right\} = \bar{L}_B. \tag{58}$$

If solutions of these equations exist, they represent the theoretical image of an experimental arrangement which allows for the simultaneous measurement of the states defined by $A$ and $B$.

The corresponding complex-valued state variable $\bar{\phi}_{AB}(\omega) = \sqrt{\bar{\rho}_{AB}(\omega)} \exp \frac{\iota}{\hbar} \bar{S}_{AB}(\omega)$ must be a simultaneous solution of the two eigenvalue equations:

$$\hat{L}_A \bar{\phi}_{AB} = a \bar{\phi}_{AB}, \qquad \hat{L}_B \bar{\phi}_{AB} = b \bar{\phi}_{AB}. \tag{59}$$

This condition leads to mathematical questions which belong to the standard repertoire of QT [36]. Eigenfunctions belonging to two different operators can only exist if these operators commute with each other. The commutation relations of the operators $\hat{L}_A$ are given by the following: [37]:

$$[\hat{L}_A, \hat{L}_B] = -\frac{\hbar}{\iota} \hat{L}_{\{A,B\}}, \tag{60}$$

where the bracket, for general operators $\hat{U}$, $\hat{V}$, is defined by $[\hat{U}, \hat{V}] = \hat{U}\hat{V} - \hat{V}\hat{U}$. The assignment $A \Rightarrow \hat{L}_A$, with $\hat{L}_A$ defined by (46), preserves the Poisson-bracket structure of the phase space observables, as discussed in detail in I. Let us recall [see Sect. 7] that a vanishing Poisson bracket $\{A, B\} = 0$ means that $A$ is invariant under the flow defined by $B$ and vice versa. The commutation relations (60) show that this invariance property is responsible for the commutability of the corresponding phase space operators.

The relations (46), (53), (60) show already a astonishing close structural similarity between classical relations, ruling the behavior of flows in phase space, and operator relations of QT. Of course, the reasoning performed here for two observables may be extended to more than two (commuting) observables, but we do not want to go into details here. Let us just mention that there is an analogy between Liouville's integrability theorem for a complete set of observables in involution [38], and the concept of a simultaneous eigenstate of a complete set of commuting operators in QT [36].

## 10 The classical counterpart of Born's rule

Let us consider the expectation value (35), where the time-dependent state $\rho(\omega, t)$ is associated with the Hamiltonian; states associated with other "stationary" observables may be used as well, but this is the most important case.

The expression for the probability density describing the measurement of a value $a$ of the observable $A(\omega)$ in the state $\rho$ at time $t$ , say $W_\rho(a, t)$, can easily be deduced from the definition (35). Let $\mathbb{R}_A \subseteq \mathbb{R}$ denote the range of possible values of the observable $A(\omega)$. In each trial, $A(\omega)$ takes a certain value $a \in \mathbb{R}_A$ corresponding to the observed outcome $\omega$. The set of all $\omega$ mapped to $a$ is the pre-image:

$$E_A(a) = \{\omega \in \Omega \,|\, A(\omega) = a\} \tag{61}$$

of $a$. These level sets provide a partition of $\Omega$. Using this fact, Eq. (35) may be written in the form:

$$\bar{A}_\rho(t) = \int \mathrm{d}a \, a \, W_\rho(a, t), \tag{62}$$

where the probability density $W_\rho(a, t)$, associated with the observable $A$, is given by the following:

$$W_\rho(a, t) = \int_{E_A(a)} \mathrm{d}\omega \, \rho(\omega, t). \tag{63}$$

In general, the subset $E_A(a)$ will not be connected, but will consist of several disjoint connected sets $E_A^{(m)}(a)$, which are referred to as connected components of $E_A(a)$. We will neglect this possibility just as we will neglect degenerate states, both in HLLK and QT.

It is interesting to rewrite the expression (35) in still another way, in terms of the operator $\hat{L}_A$. As a first step, we use the definition (47), for $A = H$ and $\alpha = t$, to rewrite (35) in the slightly more "quantum-like" form:

$$\bar{A}_\rho(t) = \int \mathrm{d}\omega\, \phi^*(\omega, t) A(\omega) \phi(\omega, t). \tag{64}$$

Note that this classical expectation value does not depend on $S$; the two terms containing $S$ cancel each other. Next, we use the definitions (46), (43), and (30) of the quantities $\hat{L}_A$, $\bar{D}_A$, and $\bar{L}_A$, to express the observable $A$ in the following way as a sum of operators and functions:

$$A = \hat{L}_A + \frac{\hbar}{\iota} \frac{\partial A}{\partial q_k} \frac{\partial}{\partial p_k} - \frac{\partial A}{\partial p_k} \left( \frac{\hbar}{\iota} \frac{\partial}{\partial q_k} - p_k \right). \tag{65}$$

Replacing $A$ in (64) with the help of this formula, the expectation value (35) takes the form:

$$\bar{A}_\rho(t) = \int \mathrm{d}\omega\, \phi^*(\omega, t) \hat{L}_A \phi(\omega, t) + \Delta \bar{A}_\rho(t). \tag{66}$$

The first term is now in perfect quantum-like form, while the second terms stems from the difference between $A$ and $\hat{L}_A$:

$$\Delta \bar{A}_\rho(t) = \int \mathrm{d}q\, \mathrm{d}p\, \phi^*(q, p, t) \left[ \frac{\hbar}{\iota} \frac{\partial A}{\partial q_k} \frac{\partial}{\partial p_k} - \frac{\partial A}{\partial p_k} \left( \frac{\hbar}{\iota} \frac{\partial}{\partial q_k} - p_k \right) \right] \phi(q, p, t). \tag{67}$$

The set of eigenfunctions of the self-adjoint operator $\hat{L}_A$ constitutes a basis in the present Koopman–von Neumann Hilbert space of square-integrable phase space functions. We assume for simplicity that the operator $\hat{L}_A$ has a discrete spectrum; its eigenvalues and orthonormalized eigenfunctions are denoted by $a_k$ and $\bar{\phi}_{A,k}(q, p)$. Expanding $\phi(q, p, t)$ and $\phi^*(q, p, t)$ in this basis and performing some well-known steps [36], we obtain the representation:

$$\bar{A}_\rho(t) = \sum_k a_k \left| (\phi, \bar{\phi}_{A,k}) \right|^2 + \Delta \bar{A}_\rho(t). \tag{68}$$

In deriving this result, we neglected again the possibility of degenerate states; appropriate generalizations can usually be carried out without problems.

We obtained two quite different representations, Eqs. (62) and (68), for the fundamental expression (35). The first expression (62) is much simpler than the second, and allows for an obvious interpretation in terms of measurement results $a$ and corresponding probabilities. Such an interpretation does not exist for the operator representation (68); the numbers $a_k$ do not belong to the experimentally verifiable output of our classical theory and a probability density describing the measurement of these numbers cannot be read off from (68).

There is a certain arbitrariness in the choice of the relevant operator, as mentioned already and as discussed in more detail in I. For example, the Koopman–von Neumann operator $\hat{D}_A$ and the real-valued state variable $\rho$ have been used to derive a classical Born-like representation for the probabilities [39]. The present classical operator representation, using $\hat{L}_A$ and complex-valued state variables, does not take the form of Born's rule [because of the term $\Delta \bar{A}_\rho(t)$] . It represents, as shown below, the classical counterpart of Born's rule, in the sense that the true quantum Born rule can be derived from it.

## 11 The transition to quantum theory

Our set of basic classical equations in operator form contains the time-dependent evolution Eq. (55), the eigenvalue Eq. (56), the commutation relations (60), and the operator form (68) for the expectation value of an observable. All these relations are purely classical, but show a remarkable structural similarity with the most important relations of QT. Such similarities may also be found in the original Koopman–von Neumann theory, where $S$ is a solution of the Liouville equation; see [40–44]. However, the present replacement of the Koopman–von Neumann operator $\hat{D}$ by the operator $\hat{L}$ makes this structural similarity much stronger.

Which modifications will lead us from these quantum-like relations to quantum relations? All fields of physics, including the complex-valued wave functions of QT, are functions of $q$, $t$ only and do not depend on the generalized momenta $p$. This seems intuitively clear, because the variables $q$, $t$ are directly measurable coordinates of our every day's "real" space–time environment. The $p's$, on the other hand, are derived quantities which cannot be measured directly. Thus, the coordinates $q$, $t$ are most fundamental (they are of course not unique, but all possible coordinate systems must be derivable from them). We may consider this as a principle of nature.

Starting from a classical ensemble of trajectories, we derived in this work a field theory with independent variables $q$, $p$, $t$, which is *in conflict* with this principle of nature. To restore this principle, we have to get rid of the coordinates $p$. A systematic way to do this, to be reported elsewhere, is to replace $p$ by a momentum field depending on $q$, $t$. However, our present quantum-like form of HLLK allows a simpler way to perform the transition, a kind of short-cut derivation of QT.

Let us rederive for completeness the simple "quantization rules", reported already in I, by considering the evolution Eq. (55) in explicit form:

$$\left[ \frac{\hbar}{\iota} \frac{\partial}{\partial t} - \frac{\hbar}{\iota} \frac{\partial H}{\partial q_k} \frac{\partial}{\partial p_k} + \frac{\partial H}{\partial p_k} \left( \frac{\hbar}{\iota} \frac{\partial}{\partial q_k} - p_k \right) + H \right] \phi = 0. \tag{69}$$

The elimination of the variables $p_k$ in this equation may be performed by assuming that $\phi$ depends not on $p_k$, and by implementing the quantization rules:

$$\frac{\partial}{\partial p_k} = 0, \quad p_k = \frac{\hbar}{\iota} \frac{\partial}{\partial q_k} \tag{70}$$

in the operator acting on $\phi(q, t)$. This step eliminates the second and third terms in the bracket of (69) and replaces the observable $H(q, p)$ by an operator $\hat{H}$, according to the general rule:

$$A \rightarrow \hat{A} = A\left( q, \frac{\hbar}{\iota} \frac{\partial}{\partial q_k} \right). \tag{71}$$

As a result, Eq. (69) becomes Schrödinger's equation. Considering the evolution equation in the form (55), the quantization process may be described in the even more compact form $\hat{L} \searrow \hat{H}$. The corresponding general rule for arbitrary observables is given by the following:

$$\hat{L}_A \searrow \hat{A}. \tag{72}$$

We are using here the terms "quantization" and "quantization rule", despite the fact that the present derivation differs considerably from the common quantization procedure.

If the same quantization rule [(70) or (72)] is applied to the eigenvalue Eq. (56) of HLLK, one obtains the eigenvalue equations of QT:

$$\hat{A} \bar{\phi}_{A,k} = a_k \bar{\phi}_{A,k}, \quad \hat{B} \bar{\phi}_{B,k} = b_k \bar{\phi}_{B,k}, \quad \text{etc}, \tag{73}$$

where the quantum operators $\hat{A}$, $\hat{B}$, ... are defined according to (71), and the eigenfunctions $\bar{\phi}_{A,k}$, $\bar{\phi}_{B,k}$, ..., associated with the eigenvalues $a_k$, $b_k$, ... of the operators $\hat{A}$, $\hat{B}$, ..., depend only on $q$.

One might ask if this simple derivation of the most fundamental differential equations of QT is an accident, but this question must be answered in the negative. As mentioned already, the transition to QT consists of two steps: A replacement of the momenta by fields, and a linearization. The reason why the simple quantization rule (70) works is the fact that the evolution Eq. (55) and the eigenvalue Eqs. (56) are already linear.

Let us turn now to the problem of several observables; it is sufficient to consider two observables. In HLLK, the condition for the simultaneous measurement of two ensemble states, defined by the observables $A$ and $B$ [see Eq. (59)], is the vanishing of the commutator of the operators $\hat{L}_A$ and $\hat{L}_B$. This vanishing is a consequence of $A$ and $B$ being in involution. Application of the quantization rule (72),

$$\{A, B\} = 0 \Rightarrow [\hat{L}_A, \hat{L}_B] = 0 \searrow [\hat{A}, \hat{B}] = 0, \tag{74}$$

leads to the well-known condition $[\hat{A}, \hat{B}] = 0$ of QT, for the existence of common eigenfunctions of the operators $\hat{A}$ and $\hat{B}$. As noted already in I, the complete structural similarity between Poisson brackets and commutator brackets, which still holds true for the transition $A(q, p) \Rightarrow \hat{L}_A$, breaks down under the transition (72) from phase space to configuration space. Thus, the transition $A \rightarrow \hat{A}$ does, in general, not preserve the Lie bracket structure [23]. This is no surprise considering the enormous reduction in the number of degrees of freedom brought about by this transition. Likewise, it is no surprise considering the fact that a single particle is different from an ensemble of particles. It is however a surprise, and an inconsistency, if one believes that the classical counterpart of a "quantum particle" is a single classical particle [45].

Let us finally turn to the quantization of the expectation value $\bar{A}_\rho(t)$ represented either in the form (66) or (68). In addition to the quantization rules (70), we have to assume now that the range of integration in all integrals is restricted to configuration space. Implementing in Eq. (66) this additional assumption as well as the quantization rules (70), we see that the term $\Delta \bar{A}_\rho(t)$ vanishes, and $\bar{A}_\rho(t)$ reduces to the quantum-mechanical expectation value of an operator $\hat{A}$ in a state $\phi$:

$$\bar{A}_\rho(t) \rightarrow (\phi, \hat{A}\phi) = \int \mathrm{d}q \, \phi^*(q, t) A\left(q, \frac{\hbar}{\imath} \frac{\partial}{\partial q_k}\right) \phi(q, t). \tag{75}$$

The quantization of $\bar{A}_\rho(t)$ in the form (68) leads in an analogous way to the quantum-mechanical representation of the expectation value of $\hat{A}$ as a weighted sum over the eigenvalues of $\hat{A}$:

$$\bar{A}_\rho(t) \rightarrow \sum_k a_k \left| (\phi, \bar{\phi}_{A,k}) \right|^2. \tag{76}$$

Here, $a_k$ and $\bar{\phi}_{A,k}$ are to be understood as eigenvalues and eigenstates of the operator $\hat{A}$ and the inner product is defined as a integral over configuration space. Both representations of $\bar{A}_\rho(t)$ are equivalent; Eq. (76) may also be obtained from (75) performing appropriate calculations in configuration space. From the representation (76), Born's rule, which says that $\left| (\phi, \bar{\phi}_{A,k}) \right|^2$ is the probability to obtain a measurement result $a_k$ of the observable $\hat{A}$ if the considered system is in the state $\phi$, can be immediately read off. The present derivation indicates that the terms "system", "observable", and "measurement result" should be interpreted in the sense of statistical ensembles and not in the sense of individual particles.

The derivation of these formulas completes the transformation from phase space to configuration space of the most important formulas of HLLK, creating thereby the most important formulas of QT. These results verify the conclusion obtained in Sect. 1, that QT must be a substructure of a probabilistic description of classical particles. The present approach provides an explanation of QT which is of unusual simplicity.

## 12 Discussion

Quantum theory in its present form was invented during the first decades of the last century. The process of invention was a combined effort of theory and experiment, adapting the theory in such a way that empirical requirements were met whenever they arose. This way of proceeding led to a number of empirically verified formal results, but gave no generally accepted interpretation, or derivation, of these results. A list of empirically verified results of QT includes the following main points:

1. Schrödinger's equation as fundamental dynamical law and eigenvalues as observable numbers.
2. The nonstandard probabilistic structure of QT—in particular non-commuting observables.
3. Born's rule—the law which tells us how to extract probabilistic predictions from the theory.
4. The minimal-coupling rule—the way interactions are formulated in QT.
5. The existence of spin—a particularly mysterious phenomenon believed to belong to QT exclusively
6. The anomalous value of the magnetic moment of the electron—a spin-related phenomenon
7. The spin–statistics connection—a spin-related multi-particle phenomenon

All these results, as well as their associated problems of interpretation, belong to the realm of non-relativistic few-particle QT. Taking into account a relativistic space–time structure or many degrees of freedom leads to more precise predictions, but does not solve the associated problems of interpretation.

The fundamental assumption underlying the present work is the old idea—put forward by Einstein [46], Born [47], and others—that the classical counterpart of QT might be a probabilistic theory. This offers—by means of the simple counting of degrees of freedom performed in Sect. 2—a unexpected possibility, namely to understand QT as a substructure of classical probabilistic physics. Planck's constant $\hbar$ was introduced as a consequence of the projection of HLLK from phase space to configuration space. Prior to this projection, $\hbar$ did not appear in any prediction of HLLK; the dependence on $\hbar$ was spurious. The appearance of a new fundamental constant $\hbar$ may be interpreted in different ways. According to a common "deterministic" metaphysics, shared by Einstein and others, it is an indication of some deeper physics. From the "indeterministic" metaphysical point of view, promoted for example by Popper [48], it realizes just an accuracy limit—indicating the elimination of the unrealistic assumption of arbitrary high accuracy.

In the present work, we showed, by deriving QT from HLLK, that QT is a substructure of an extended version of probabilistic classical physics. The very success of such a calculation represents, of course, a strong argument in favor of Einstein's ensemble interpretation of QT [1,46,49]. As regards questions of interpretation, we do not want to go into details here; let us just note that our results do not support concepts like "completeness" or "nonlocality" of QT [50]. Summarizing, we could say "Einstein was right", quoting the title of a recent careful analysis [51] of Bell's inequality.

The present derivation is based on a few general and reasonable assumptions, such as the superiority of a probabilistic description as compared to a deterministic description, and the superiority of the space–time coordinates as compared to the space–momentum–time coordinates. We were able to derive some crucial features of QT, namely those corresponding to points 1–3 in the above list. While there are still open points, it seems that such a coherent derivation of fundamental concepts of QT has never been obtained before. We expect, therefore, that a complete derivation of QT, including all points in the above list, is possible. However, it is clear that such a complete derivation of QT requires a more detailed analysis of the steps leading to QT than given here.

## 13 Concluding remarks

One of the most fundamental conclusions of the present work is the following: not classical physics "emerges" from QT in the limit $\hbar \rightarrow 0$, but the inverse is true; QT emerges from the classical theory HLLK. The new quantum constant $\hbar$ appears as a consequence of a projection of HLLK to configuration space. Using a more philosophical terminology, we might say that the otherwise powerful principle of reductionism is not useful for clarifying the relation between classical physics and QT.

The question why Schrödinger's equation is complex is sometimes considered as crucial for the understanding of QT. From the present point of view, the complexity is a consequence of the requirement of linearity of the basic equation. All fundamental field equations of physics (equations *not* containing any additional material parameters) must be linear. Otherwise, they would be unable to describe a large number of individual events—simply because almost all solutions of nonlinear equations become singular after a sufficient long time. A linear theory containing two real-valued dynamical variables can only be constructed if a complex-valued state variable is introduced.

In the present theory with two real variables, we derived as a special case, for $N = 1$, Schrödinger's equation describing single particle ensembles of "massive spinless particles". According to present-day's experimental evidence, such objects do not exist; all structure-less massive particles found in nature have spin one-half. This indicates that the present derivation contains an error. This error is implicitly contained in the present somewhat crude projection to configuration space. A more careful projection contains as a first step a replacement of the momenta by fields depending on $q$, $t$. If this replacement is performed in a correct way, one obtains for a single particle (ensemble) a theory with four real variables, which means that a single particle has spin one-half. Details will be reported in forthcoming work.

**Acknowledgements** Open access funding provided by Johannes Kepler University Linz.

**Compliance with ethical standards**

**Conflicts of interest** The author states that there is no conflict of interest.